# Risk Sensitive Path Integral Control

**Bart van den Broek**     **Wim Wiegerinck**     **Bert Kappen**

SNN, Radboud University Nijmegen
6525 EZ Nijmegen, The Netherlands
{b.vandenbroek,w.wiegerinck,b.kappen}@science.ru.nl

## Abstract

Recently path integral methods have been developed for stochastic optimal control for a wide class of models with non-linear dynamics in continuous space-time. Path integral methods find the control that minimizes the expected cost-to-go. In this paper we show that under the same assumptions, path integral methods generalize directly to risk sensitive stochastic optimal control. Here the method minimizes in expectation an exponentially weighted cost-to-go. Depending on the exponential weight, risk seeking or risk averse behaviour is obtained. We demonstrate the approach on risk sensitive stochastic optimal control problems beyond the linear-quadratic case, showing the intricate interaction of multi-modal control with risk sensitivity.

## 1  Introduction

The objective in conventional stochastic optimal control is to minimize an expected cost-to-go [10]. Risk sensitive optimal control generalizes this objective by minimizing an expected exponentiated cost-to-go. Depending on its risk parameter, expected exponentiated cost-to-go puts more emphasis on the mode of the distribution of the cost-to-go, or on its tail, and in that way allows for a modelling of more risk seeking or risk averse behaviour. The conventional optimal control can be viewed as a special case of risk sensitive optimal control with a risk neutral parameter.

Risk sensitive control was first considered by Howard and Matheson [12] in discrete space, and by Jacobson [13] in continuous space in the LEQG (Linear Exponential Quadratic Gaussian) problem, which is the risk sensitive analogue of the linear quadratic Gaussian (LQG) problem. Relations with other fields such as differential games [13, 9] and robust control [11, 6] have initiated a lot of interest for risk sensitive control.

The dynamic programming (DP) principle by Bellman [2] provides a well-known approach to a global solution in stochastic optimal control. In the continuous time and state setting that we will consider, it follows from the DP principle that the solution to the control problem satisfies the so-called Hamilton-Jacobi-Bellman (HJB) equation, which is a second order non-linear partial differential equation [10, 18, 3]. If the dynamics is linear and the cost is quadratic in both state and control, the HJB equation can be solved exactly, both for LQG and LEQG.

Recently, a path integral formalism has been developed to solve the HJB equation. This formalism is applicable if (1) both the noise and the control are additive to the (nonlinear) dynamics, (2) the increment in cost is quadratic in the control, and (3) the noise satisfies certain additional conditions. When these conditions are met, it can be shown that the nonlinear HJB equation can be transformed into a linear partial differential equation, which can be solved by forward stochastic integration of a diffusion process [16, 17]. This formalism contains LQG control as a special case. Recently, path integral optimal control has been identified as a member of a richer framework of control models, in which the cost function is written as a KL-divergence [20].

An interesting observation in [16, 17] is the phenomenon of symmetry breaking in multi-modal systems, i.e, in problems where several local minima of the cost co-exist. This symmetry breaking manifests itself as a delayed choice, keeping options open and using the fact that the noise may help to come closer to one of the options at no additional cost.

Path integral methods have been applied to optimal control of collaborative multi-agent systems [23, 24]. The optimal control is written as a superposition of single-agent to single-target controls, and the multi-agent control problem is mapped to a graphical infer-ence problem. Similar superposition principles have been used in applications, e.g. the control of character animation [5]. In general, the path integral controls are intractable, and approximate inference methods must be applied [22].

In this paper we show how path integral control generalizes to risk sensitive control problems. The required conditions to apply path integral control in the risk sensitive case are the same as those in the risk neutral setting. As a consequence, the characteristics of path integral control, such as superposition of controls, symmetry breaking and approximate inference, carry over to the setting of risk sensitive control. Furthermore, we will make use of the path integral solutions to obtain insight in the consequence of risk sensitive control, and in particular interpret risk sensitive optimal control as emergent behaviour of an agent with an optimistic or pessimistic attitude.

We start with a review of risk sensitive control in section 2. The novel generalization of path integral control to risk sensitive control problems is presented in section 3. In sections 4 and 5 we give a more informal account on the behaviour of risk sensitive control problems with path integral control. We finish with a discussion in section 6.

## 2 RISK SENSITIVE STOCHASTIC OPTIMAL CONTROL

In this section we review risk sensitive stochastic optimal control. Risk neutral stochastic optimal control is covered as a special case. Details can be found in e.g. [3, 8, 14].

We consider a stochastic process $X^u$ in $\mathbb{R}^d$ satisfying the controlled dynamics

$$dX_t^u = b(t, X_t^u)dt + B(t, X_t^u)\Big(u(t, X_t^u)dt + \sigma dW_t\Big), \quad (1)$$

where $b$ is the $\mathbb{R}^d$-valued autonomous dynamics, $B$ is a full rank $d \times k$ matrix-valued function, $k \leq d$, $u$ is a $\mathbb{R}^k$-valued control, $\sigma$ is a constant full rank $k \times k$ matrix, and $W$ is a Wiener process in $\mathbb{R}^k$ which models the noise in the system.

The performance of the system is evaluated by the cost function

$$C^u(t, X^u) = \phi(X_T^u) + \int_t^T L^u(s, X_s^u)ds,$$

where $\phi$ is the end cost and $L^u$ is an instantaneous cost. For any $\theta \in \mathbb{R}$ we define a value function

$$J_\theta^u(t, x) = \begin{cases} \mathbb{E}_{t,x}\big[C^u(t, X^u)\big] & \text{if } \theta = 0, \\ \frac{1}{\theta} \log \mathbb{E}_{t,x}\big[\exp\big(\theta\, C^u(t, X^u)\big)\big] & \text{otherwise,} \end{cases}$$

where $\mathbb{E}_{t,x}$ denotes expectation over all realisations of the dynamics starting in $x$ at time $t$, and we define an optimal value function $J_\theta$ by

$$J_\theta(t, x) = \inf_u J_\theta^u(t, x).$$

Note that the case $\theta = 0$ is that of risk neutral control. A control $u^*$ which satisfies $J_\theta^{u^*} = J_\theta$ is called optimal.

Around $\theta = 0$, using series expansions of exp and log, the value function satisfies

$$J_\theta^u(t, x) = \mathbb{E}_{t,x}\big[C^u(t, X^u)\big]$$
$$+ \tfrac{\theta}{2}\Big(\mathbb{E}_{t,x}\big[C^u(t, X^u)^2\big] - \mathbb{E}_{t,x}\big[C^u(t, X^u)\big]^2\Big) + O(\theta^2).$$

We observe that in case of a negative $\theta$ it is favorable to have a large variation in cost, and this is interpreted as a risk seeking behaviour. A positive $\theta$ on the other hand corresponds to a risk averse behaviour.

### 2.1 RISK MONOTONICITY

In the proposition below we describe how the value function behaves as a function of $\theta$.

**Proposition 2.1** *For any fixed time $t$, state $x$, and control $u$, the mapping*

$$\mathbb{R} \ni \theta \mapsto J_\theta^u(t, x) \in [-\infty, \infty]$$

*is constant if and only if the cost $C^u(t, X^u)$ is constant with probability one, and it is strictly increasing otherwise. Its limits at $-\infty$ and $\infty$ are given by the respective extremal costs, that is,*

$$\lim_{\theta \to -\infty} J_\theta^u(t, x) = \sup\{a \in \mathbb{R} : \mathbb{P}_{t,x}(C^u(t, X^u) \geq a) = 1\},$$
$$\lim_{\theta \to \infty} J_\theta^u(t, x) = \inf\{a \in \mathbb{R} : \mathbb{P}_{t,x}(C^u(t, X^u) \leq a) = 1\}.$$

*Proof:* By Jensen's inequality, $\mathbb{E}[|Y|^a] \geq \mathbb{E}[|Y|]^a$ for any $a > 1$ and any random variable $Y$, where equality holds if and only if $Y$ is constant with probability one. It follows that for any $\theta_1$ and $\theta_2$ satisfying $0 < \theta_1 < \theta_2$,

$$\mathbb{E}\big[|Y|^{\theta_2}\big]^{1/\theta_2} \geq \Big(\mathbb{E}\big[|Y|^{\theta_1}\big]^{\theta_2/\theta_1}\Big)^{1/\theta_2} = \mathbb{E}\big[|Y|^{\theta_1}\big]^{1/\theta_1}.$$

Choosing $Y = \exp(C^u(t, X^u))$, we find that the mapping $\theta \mapsto J_\theta^u(t, x)$ is constant on $(0, \infty)$ if and only if $C^u(t, X^u)$ is constant with probability one, otherwise it is strictly increasing. We can extend this from $\theta \in (0, \infty)$ to $\theta \in \mathbb{R}$ by the fact that $J_\theta^u(t, x)$ satisfies

$$J_{-\theta}^u(t, x) = -\log \mathbb{E}_{t,x}\big[|\exp(-C^u(t, X^u))|^\theta\big]^{1/\theta}$$

for any $\theta \in (0, \infty)$, and that $\theta \mapsto J_\theta^u(t, x)$ is continuous in $\theta = 0$.

In the limit $\theta \to \infty$ we have

$$\lim_{\theta \to \infty} J^u_\theta(t, x) = \log \| \exp(C^u(t, X^u)) \|_\infty$$
$$= \log \inf \{ a \in \mathbb{R} : \mathbb{P}_{t,x}\big( \exp(C^u(t, X^u)) \leq a \big) = 1 \}$$
$$= \inf \{ a \in \mathbb{R} : \mathbb{P}_{t,x}\big( C^u(t, X^u) \leq a \big) = 1 \},$$

and in the limit $\theta \to -\infty$ we have

$$\lim_{\theta \to -\infty} J^u_\theta(t, x) = -\log \| \exp(-C^u(t, X^u)) \|_\infty$$
$$= -\log \inf \{ a \in \mathbb{R} : \mathbb{P}_{t,x}\big( \exp(-C^u(t, X^u)) \leq a \big) = 1 \}$$
$$= \sup \{ a \in \mathbb{R} : \mathbb{P}_{t,x}\big( C^u(t, X^u) \geq a \big) = 1 \}.$$

This finishes the proof. □

## 2.2 THE HJB EQUATION

We give a formal derivation of the HJB equation for risk sensitive stochastic optimal control problems in continuous space and time and with a finite horizon.

By the dynamic programming principle, the optimal cost function $J_\theta$ satisfies

$$J_\theta(t,x)$$
$$= \inf_u \tfrac{1}{\theta} \log \mathbb{E}_{t,x} \left[ \exp\left( \theta C^u(r, X^u_r) + \theta \int_t^r L^u(s, X^u_s) ds \right) \right]$$
$$= \inf_u \tfrac{1}{\theta} \log \mathbb{E}_{t,x} \left[ \exp\left( \theta J^u_\theta(r, X^u_r) + \theta \int_t^r L^u(s, X^u_s) ds \right) \right]$$
$$= \inf_u \tfrac{1}{\theta} \log \mathbb{E}_{t,x} \left[ \exp\left( \theta J_\theta(r, X^u_r) + \theta \int_t^r L^u(s, X^u_s) ds \right) \right]$$
(2)

for any time $r$, $t \leq r \leq T$. The first two identities follow directly from the definitions. We define a function $\mathcal{E}_\theta$ by

$$\mathcal{E}_\theta(t, x) = \exp(\theta J_\theta(t, x)).$$

By Itô's chain rule for stochastic processes [10], $\mathcal{E}_\theta(r, X^u_r)$ satisfies

$$\mathcal{E}_\theta(r, X^u_r) = \mathcal{E}_\theta(t, x) + \int_t^r \left( \frac{\partial}{\partial s} + \mathscr{A}^u \right) \mathcal{E}_\theta(s, X^u_s) ds$$
$$+ \int_t^r \left( \frac{\partial}{\partial x} \mathcal{E}_\theta(s, X^u_s) \right)^\top B(s, X^u_s) \sigma dW_s,$$

where $\mathscr{A}^u$ is the differential operator

$$\mathscr{A}^u = \sum_{i=1}^d (b + Bu)_i \frac{\partial}{\partial x_i} + \tfrac{1}{2} \sum_{i,j=1}^d (B\sigma\sigma^\top B^\top)_{ij} \frac{\partial^2}{\partial x_i \partial x_j}.$$

In case of zero control we simply write $\mathscr{A}$ for $\mathscr{A}^0$. We insert this expression for $\mathcal{E}_\theta(r, X^u_r)$ in equation (2). If $\theta > 0$, then we can drop the $\tfrac{1}{\theta}\log$, and we find

$$0 = \inf_u \mathbb{E}_{t,x} \left[ \mathcal{E}_\theta(t, x) \left( \exp\left( \theta \int_t^r L^u(s, X^u_s) ds \right) - 1 \right) \right.$$
$$+ \exp\left( \theta \int_t^r L^u(s, X^u_s) ds \right) \int_t^r \left( \frac{\partial}{\partial s} + \mathscr{A}^u \right) \mathcal{E}_\theta(s, X^u_s) ds$$
$$+ \exp\left( \theta \int_t^r L^u(s, X^u_s) ds \right)$$
$$\left. \int_t^r \left( \frac{\partial}{\partial x} \mathcal{E}_\theta(s, X^u_s) \right)^\top B(s, X^u_s) \sigma dW_s \right].$$

Dividing by $r - t$, and taking the limit $r \downarrow t$ yields

$$0 = \inf_u \left( \frac{\partial}{\partial t} + \mathscr{A}^u + \theta L^u \right) \mathcal{E}_\theta(t, x)$$
$$= \inf_u \theta \mathcal{E}_\theta(t,x) \left( \frac{\partial J_\theta}{\partial t} + \mathscr{A}^u J_\theta + L^u \right.$$
$$\left. + \tfrac{1}{2} \theta \left\| \sigma^\top B^\top \frac{\partial J_\theta}{\partial x} \right\|^2 \right)(t,x).$$

Dividing by $\theta \mathcal{E}_\theta(t, x)$, we find the HJB equation

$$0 = \inf_u \left( \frac{\partial J_\theta}{\partial t} + \mathscr{A}^u J_\theta + L^u + \tfrac{1}{2}\theta \left\| \sigma^\top B^\top \frac{\partial J_\theta}{\partial x} \right\|^2 \right)(t,x). \quad (3)$$

In case $\theta < 0$, the HJB equation (3) also holds, and it is derived in a similar way.

## 3 RISK SENSITIVE PATH INTEGRAL CONTROL

The novel combination of path integral control and risk sensitive control is the subject of the present section. It generalizes risk neutral path integral control which corresponds to the case $\theta = 0$. We consider instantaneous control cost functions $L^u$ of the form

$$L^u(s, x) = \tfrac{1}{2} \| Ru(s, x) \|^2 + V(s, x),$$

and we assume that

$$\sigma \sigma^\top = \lambda_0 (R^\top R)^{-1} \quad (4)$$

for some $\lambda_0 \in \mathbb{R}$. Substitution of $L^u$ in the HJB equation (3) yields

$$0 = \inf_u \left( \frac{\partial J_\theta}{\partial t} + \mathscr{A}^u J_\theta + \tfrac{1}{2}\|Ru\|^2 + V \right.$$
$$\left. + \tfrac{1}{2}\theta \left\| \sigma^\top B^\top \frac{\partial J_\theta}{\partial x} \right\|^2 \right)(t,x)$$
$$= \left( \frac{\partial J_\theta}{\partial t} + \mathscr{A} J_\theta + V \right.$$
$$\left. + \tfrac{1}{2}\left( \frac{\partial J_\theta}{\partial x} \right)^\top B \big( \theta \sigma \sigma^\top - (R^\top R)^{-1} \big) B^\top \left( \frac{\partial J_\theta}{\partial x} \right) \right)(t,x)$$

with the optimal control given by

$$u_\theta^*(t,x) = -(R^\top R)^{-1} B^\top(t,x) \frac{\partial J_\theta(t,x)}{\partial x}.$$

If $\theta = \lambda_0^{-1}$ then the HJB equation is linear, and its solution is given by the path integral

$$J_\theta(t,x) = \mathbb{E}_{t,x}\left[\phi(X_T) + \int_t^T V(s,X_s)ds\right] \quad (5)$$

due to the Feynman-Kac formula [7, 15], where $X$ satisfies the dynamics (1) without control. Otherwise, we define a logarithmic transformation of $J_\theta$ by

$$J_\theta = -\lambda_\theta \log Z_\theta, \qquad \lambda_\theta = \frac{\lambda_0}{1-\lambda_0\theta}.$$

Substituting this expression for $J_\theta$ in the HJB equation, and subsequently dividing by $-\lambda_\theta Z_\theta^{-1}$, yields

$$0 = \left(\frac{\partial Z_\theta}{\partial t} + \mathscr{A} Z_\theta - \frac{1}{\lambda_\theta} V Z_\theta \right.$$
$$\left. + Z_\theta^{-1} \tfrac{1}{2}\left(\frac{\partial Z_\theta}{\partial x}\right)^\top BMB^\top\left(\frac{\partial Z_\theta}{\partial x}\right)\right)(t,x),$$

where

$$M = \lambda_\theta(R^\top R)^{-1} - (1+\lambda_\theta\theta)\sigma\sigma^\top.$$

The term proportional to $Z_\theta^{-1}$ vanishes after substituting $\lambda_\theta = \lambda_0(1-\lambda_0\theta)^{-1}$, and the resulting HJB equation is linear. The solution of the linear HJB equation is given by the path integral

$$Z_\theta(t,x) = \mathbb{E}_{t,x}\left[\exp\left(-\tfrac{1}{\lambda_\theta}\phi(X_T) - \tfrac{1}{\lambda_\theta}\int_t^T V(s,X_s)ds\right)\right]. \quad (6)$$

The optimal value function satisfies

$$J_\theta(t,x) = \log \mathbb{E}_{t,x}\left[\exp\left(C(t,X)\right)^{-1/\lambda_\theta}\right]^{-\lambda_\theta},$$

where $C$ is the cost function under zero control, with the limit case at $\theta = \lambda_0^{-1}$ given by equation (5). Since $-\frac{1}{\lambda_\theta} = \theta - \frac{1}{\lambda_0}$, we find in a way similar as in Proposition 2.1 that the optimal value function $J_\theta$ is constant as a function of $\theta$ if and only if the cost $C(t,X)$ under zero control is constant with probability one, and it is strictly increasing otherwise.

In the limit $\theta \to -\infty$ (i.e., $\lambda_\theta^{-1} \to \infty$) we find

$$\lim_{\theta \to -\infty} J_\theta(t,x) = \inf_X C(t,X)$$

and in the limit $\theta \to \infty$ (i.e., $\lambda_\theta^{-1} \to -\infty$) we find

$$\lim_{\theta \to \infty} J_\theta(t,x) = \sup_X C(t,X)$$

In particular, this shows that if either of these limits is finite, $J_\theta(t,x)$ is independent of $x$ and the optimal control will be zero.

One could argue that $\theta \to -\infty$ is the extreme optimistic limit. In this limit, $J_\theta = \inf_X C(t,X)$, which is the cost under zero control following the most optimistic path due to the noise leading to the lowest possible cost. In this extreme optimistic view, the noise will lead the agent to the most optimal cost. There is no need for control, since this will only lead to additional costs. On the other hand, $\theta \to \infty$ is the extreme pessimistic limit. $J_\theta = \sup_X C(t,X)$, which is the cost under zero control following the most pessimistic path due to the noise leading to the highest possible cost. According to this extreme pessimistic view, any additional control is pointless, since the noise realization will be such that the worst path will be realized anyway.

## 4 LINEAR EXPONENTIAL QUADRATIC GAUSSIAN

The running example is a one-dimensional system with linear dynamics, $b=0$, and zero path-costs $V(t,x)=0$, $B=1$, $R$ and $\sigma$ are proportional to the identity and are considered as scalars. In this section we furthermore take a quadratic end cost around a target $\mu$,

$$\phi(x) = \frac{\alpha^2}{2}|x-\mu|^2.$$

Under these assumptions, in particular the quadratic end-costs, the optimal control can be computed using LEQG (Linear Exponential Quadratic Gaussian) theory [4]. The results in this section are mainly to illustrate the path integral approach to risk sensitive control.

Since $V(t,x) = 0$, the expectation $\mathbb{E}_{t,x}$ can be computed by a convolution with the transition probability from initial state to end state, $\rho(y,T|x,t)$, that follows from the zero-control dynamics. In this example, this transition probability is

$$\rho(y,T|x,t) = \frac{1}{\sqrt{2\pi\sigma^2(T-t)}}\exp\left(-\frac{(y-x)^2}{2\sigma^2(T-t)}\right),$$

so the path integral $Z_\theta(t,x)$ follows from a Gaussian convolution with $e^{-\alpha^2|x-\mu|^2/2\lambda_\theta}$, yielding up to a finite prefactor

$$Z_\theta(t,x) \propto \int_{-\infty}^\infty \exp\left(-\frac{(y-x)^2}{2\sigma^2(T-t)} - \frac{\alpha^2(y-\mu)^2}{2\lambda_\theta}\right)dy.$$

If

$$\theta < \frac{1}{\alpha^2\sigma^2(T-t)} + \frac{1}{\sigma^2 R^2} \quad (7)$$

is not satisfied, then the integral blows up and $Z_\theta(t, x)$ is infinite, otherwise, the integral is well defined, and the resulting optimal control can be computed, yielding

$$u_\theta^*(t, x) = \frac{\alpha^2(\mu - x)}{R^2 + (T-t)\alpha^2(1 - \sigma^2 R^2 \theta)}.$$

Generally speaking, we see that agents with larger $\theta$ have a larger control: $\theta_1 < \theta_2 \to |u_{\theta_1}^*(t,x)| < |u_{\theta_2}^*(t,x)|$. Furthermore, we see that if $\theta < \frac{1}{\sigma^2 R^2}$, so $\lambda_\theta^{-1} > 0$, the amplitude of the control $|u_\theta^*(t,x)|$ for fixed $x$ is zero in the limit of large time horizons $T-t$, and it increases over time. If $\theta = \frac{1}{\sigma^2 R^2}$, so $\lambda_\theta^{-1} = 0$, the amplitude of the control $|u_\theta^*(t,x)|$ for fixed $x$ is constant in time. On the other hand, if $\theta > \frac{1}{\sigma^2 R^2}$, so $\lambda_\theta^{-1} > 0$, the amplitude of the control $|u_\theta^*(t,x)|$ is undefined if the time horizon is larger than $\sigma^2 \alpha^2/(\theta - \lambda_\theta^{-1})$, it is infinite at exactly this limit and decreases over time if the horizon is smaller than this limit. With larger horizon times, the expectation of reaching infinite costs in the tails of the quadratic function due to the noise blows up. These results are well known from LEQG control theory [4].

# 5 PIECEWISE CONSTANT END-COSTS

In this section we again consider control problems with zero path costs $V(t,x) = 0$. We consider piecewise constant end-costs $\phi(x)$. Thus, these control problems are not of the LEQG kind. We assume $B = I$ for simplicity. We will show that the control with given $\theta$ can be expressed as a weighted sum of controls with $\theta = 0$, similar to the approach in [23, 24] and later more general in [21]. This expression will allow us to analyse the different behaviours for agents with different $\theta$'s.

## 5.1 A SINGLE REGION

First we consider the standard $\theta = 0$ case where the agent starts in state $x$ at time $t$, and the end-cost is finite in a certain region $S$ and infinite otherwise,

$$\phi(x) = \begin{cases} c & \text{if } x \in S \\ \infty & \text{otherwise.} \end{cases} \quad (8)$$

The path integral $Z_\theta(t,x)$, also interpreted as a partition function, is then given by $Z_\theta(t,x) = \exp(-c/\lambda_0)l(t,x|S)$, with

$$l(t,x|S) = \int_S \rho(y,T|x,t)dy \quad (9)$$

with $\rho(y,T|x,t)$ the transition probability from initial state to end state according to the zero-control dynamics. The $\theta = 0$ optimal control with this cost then is

$$u_0^*(t,x|S) = \frac{\lambda_0(R^\top R)^{-1}}{l(t,x|S)} \frac{\partial l(t,x|S)}{\partial x} \quad (10)$$

which is independent of the value of $c$.

## 5.2 WEIGHTED SUM OF CONTROLS

Now we consider nonzero $\theta$ and an end-cost function

$$\phi(x) = c_i \quad \text{if } x \in S_i, \quad i = 0, \ldots, M \quad (11)$$

in which the $S_i$ are non-overlapping sets covering the whole state space.

The partition function according to (6) results in

$$Z_\theta(t,x) = \sum_{i=0}^M \exp(-c_i/\lambda_\theta)l(t,x|S_i).$$

By taking the derivative of $\log Z_\theta(t,x)$, the optimal control follows directly and can be written as a weighted sum of $\theta = 0$ optimal controls:

$$u_\theta^*(t,x) = \frac{\lambda_\theta}{\lambda_0} \sum_{i=0}^M w_i(t,x) u_0^*(t,x|S_i) ,$$

with weights

$$w_i(t,x) = \frac{\exp(-c_i/\lambda_\theta)l(t,x|S_i)}{\sum_{i=0}^M \exp(-c_i/\lambda_\theta)l(t,x|S_i)}.$$

## 5.3 TARGETS AND THREATS

A special case is where there are $M$ bounded regions $S_i$, and an unbounded remaining set $S_0 = \mathbb{R}^d \setminus \cup S_i$. We may set $c_0 = 0$, since a global additive constant to the end-cost will not affect the control. Then there are two types of regions, those with negative costs, $c_i < 0$, and those with positive costs, $c_i > 0$. The former are interpreted as targets, since it is favorable to arrive there, and the latter as threats, since these are better avoided.

We rewrite the partition function as

$$Z_\theta(t,x) = 1 + \sum_{i=1}^M (\exp(-c_i/\lambda_\theta) - 1)l(t,x|S_i).$$

so the optimal control is now

$$u_\theta^*(t,x) = \frac{\lambda_\theta}{\lambda_0} \sum_{i=1}^M w_i(t,x) u_0^*(t,x|S_i) , \quad (12)$$

with weights

$$w_i(t,x) = \frac{(\exp(-c_i/\lambda_\theta) - 1)l(t,x|S_i)}{1 + \sum_{i=1}^M (\exp(-c_i/\lambda_\theta) - 1)l(t,x|S_i)}.$$

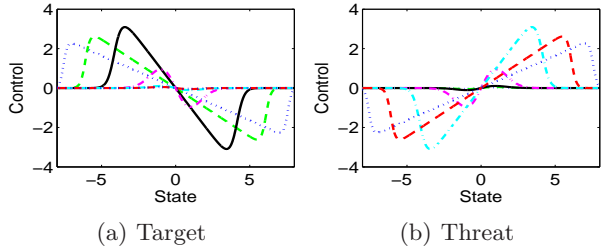

Figure 1: The control as a function of the state in case of an end target (a) or end threat (b), with $\lambda_\theta = -\frac{1}{3}$ $(\cdots)$, $-\frac{1}{2}$ $(--)$, $-1$ $(-\cdot-)$, $\pm\infty$ $(-\cdot-)$, $1$ $(-)$, $\frac{1}{2}$ $(--)$, $\frac{1}{3}$ $(\cdots)$.

Here we note that if $\lambda_\theta \gg 0$, the control is dominated by the targets, due to the weights $w_i$ with large factors $\exp(-c_i/\lambda_\theta)$. In a similar way, if $\lambda_\theta \ll 0$, the control is dominated by the threats. Due to the prefactor $\lambda_\theta/\lambda_0$ in (12), which is negative in this case, the control is directed away from the threats.

This is illustrated in the first example, see Figure 1. We consider the end cost given by equation (11) with a single small region $S_1$ around $x = 0$. Depending on the sign of $c$, this models a single target ($c < 0$) or a single threat ($c > 0$) in $S_1$. We assume $b(t, x) = 0$ so that $\rho$ is Gaussian and $l(t, x|S_1)$ can be expressed in closed form using error functions. In the figure, the controls as a function of $x$ for fixed $t$ are plotted for different $\theta$'s, for both target and threat.

In the figure, we see indeed only significant controls in combinations of positive $\lambda_\theta^{-1}$ and targets on the one hand and negative $\lambda_\theta^{-1}$ and threats on the other hand. The case with $\lambda_\theta^{-1} = 0$ is somewhere in between. Furthermore, we see that the nonzero control has a bounded support, that increases if $|\lambda_\theta^{-1}|$ increases. This can be understood from the fact that the control is only significantly nonzero if the weight $w_1$ is significantly nonzero, i.e., if the product of $|(\exp(-c/\lambda_\theta) - 1)|$ and $l(t, x|S_1)$ is significantly big. For the factor $l(t, x|S_1)$, which is independent of $\theta$, this means that the agent must get sufficiently close to the target/threat. If not, control cost towards the target is too expensive compared to the reward, or the probability to hit the threat is so small that a significant control is not needed. With larger values of $-c/\lambda_\theta$, the domain for which $|(\exp(-c/\lambda_\theta) - 1)|l(t, x|S_1)$ and hence $u_\theta^*(t, x)$ is significantly nonzero is larger, as can be seen in the figure. On the other hand, with larger $|\lambda_\theta^{-1}|$ the prefactor $\lambda_\theta/\lambda_0$ in (12) will be smaller, so that in the regime where $|(\exp(-c/\lambda_\theta) - 1)|l(t, x|S_1)$ is large, the control decreases with larger $|\lambda_\theta^{-1}|$, as can be seen in the figure.

The second example considers the phenomenon of symmetry breaking with different values of $\theta$ and $c$. In the case of $\theta = 0$ and two targets, the phenomenon of symmetry breaking in the optimal control, leading to a delayed choice, is described using the path integral formalism in [16, 17]. Here we show that for other values of $\theta$, such that $c\lambda_\theta^{-1} \gg 0$ the symmetry breaking with two targets remains essentially the same. See Figure 2. There are two target regions of size $\epsilon$, located at $-1$ and $+1$. The left figure shows that if $T - t = 1$, the optimal control is towards the middle for all values of $x$. The right figure shows that there is symmetry breaking at time $T - t = 0.5$, i.e., depending on the state $x$, the agent makes a choice towards which target it steers. Furthermore, we see that the point of symmetry breaking is independent of $\theta$, although the magnitude of control does depend on $\theta$. For instance, we see that if $\theta$ is such that $\lambda_\theta^{-1} < 0$, the control is about zero. This latter phenomenon can be understood from the fact that with such $\theta$, the control is much less sensitive to targets, as we have discussed earlier.

The symmetry breaking for such $\theta$ where $\lambda_\theta^{-1}$ is large can be understood as follows. Consider again $b = 0$, so that $\rho$ is Gaussian. We model two targets at $\mu_1 = 1$ and $\mu_2 = -1$ of infinitesimal width $\epsilon$. Furthermore, assume that $-c\lambda_\theta^{-1}$ is sufficiently large, so that the factor $\exp(-c\lambda_\theta^{-1})l(t, x|[\mu_i - \epsilon/2, \mu_i + \epsilon/2]) \approx \alpha\delta(y - \mu_i)\rho(y, T|x, t)$, with $\alpha > 0$ a global constant. In this limit, the optimal control is

$$u_\theta^*(t, x) = \frac{\lambda_\theta}{\lambda_0(T-t)}\left(\tanh\left(\frac{x}{\sigma^2(T-t)}\right) - x\right).$$

Regardless the precise value $\lambda_\theta^{-1} \gg 0$, the control displays a symmetry breaking at $T - t = 1/\sigma^2$. For earlier times, it is best to steer towards $x = 0$ (between the targets) and delay the choice of which target to aim for until later. The reason why this is optimal is that from that position the expected diffusion alone is likely to reach any of the slits without control (although it is not clear yet which target). Only sufficiently late in time should one make a choice and steer towards one of the targets, instead of towards the middle. A more careful analysis without using delta functions shows that the phenomenon of symmetry breaking is also present with finite valued targets. Also in that case, the time of symmetry breaking is independent of $\theta$.

For threats, i.e., if the regions around $\mu_i$ have positive costs and are to be avoided, a similar but reversed phenomenon occurs. See Figure 3. For $T - t = 1$, the optimal control is outwards the middle for all values of $x$, trying for a global escape away from both threats (left figure). For later times, if the agent is somewhere in the middle, an escape passing one of the threats would be too risky, and the agent decides to remain in the middle of the two threats. Again, the point of symmetry breaking does not depend on $\theta$, but the magnitude of the control does. In particular, agents with $\lambda_\theta^{-1} < 0$ now have big controls, whereas $\lambda_\theta^{-1} > 0$ seems hardly to care, as we discussed earlier.

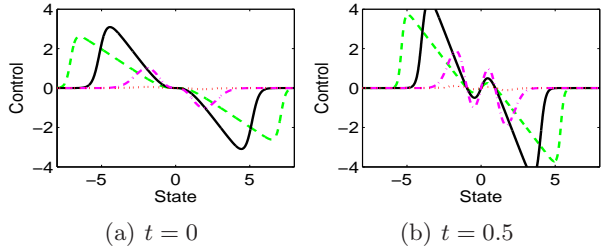

(a) $t = 0$

(b) $t = 0.5$

Figure 2: The control as a function of the state in case of two end targets with $\lambda_\theta = -\frac{1}{2}$ ($\cdots$), $\pm\infty$ ($-\cdot-$), $1$ ($-$), $\frac{1}{2}$ ($--$). There is no symmetry breaking at time $t = 0$ (a), but there is at time $t = 0.5$ (b).

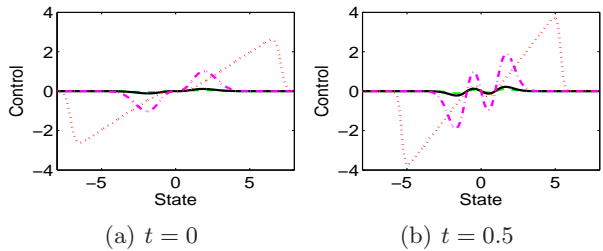

(a) $t = 0$

(b) $t = 0.5$

Figure 3: The control as a function of the state in case of two end threats with $\lambda_\theta = -\frac{1}{2}$ ($\cdots$), $\pm\infty$ ($-\cdot-$), $1$ ($-$), $\frac{1}{2}$ ($--$). There is no symmetry breaking at time $t = 0$ (a), but there is at time $t = 0.5$ (b).

A third case that we consider is a target and an adjacent threat: $S_1 = \{-0.1 < x < 0\}$, $c_1 = -10$ and $S_2 = \{0 < x < 0.1\}$, $c_2 = 10$. We did runs for $\theta = -1, 0, 1, 3$. For each $\theta$, we did 1000 runs. Each run started at time $t = 0$ and position $x = 0$. End time is $T = 1$. For each run we monitored the total cost $C$ (i.e. $C$ is control costs plus end cost). Histograms of the log-probability of the cost are plotted in Figure 4. We see that with larger $\theta$, the mode of the distribution shifts to higher costs. On the other hand, the tails of the distribution at the high cost end are thinner with larger $\theta$. This is what is to be expected: small $\theta$ is more greedy, aiming at low cost, however at the expense of some outliers with high costs. A larger $\theta$ is more cautious, reducing the probability of costly outliers.

# 6 DISCUSSION

We showed that path integral control is applicable to risk sensitive control problems. In particular, we showed that the path integral formalism is applicable whenever it is applicable to risk neutral ($\theta = 0$) problems, provided that the path integral does not diverge. Thus the class of stochastic optimal control problems that can be solved exactly is enlarged.

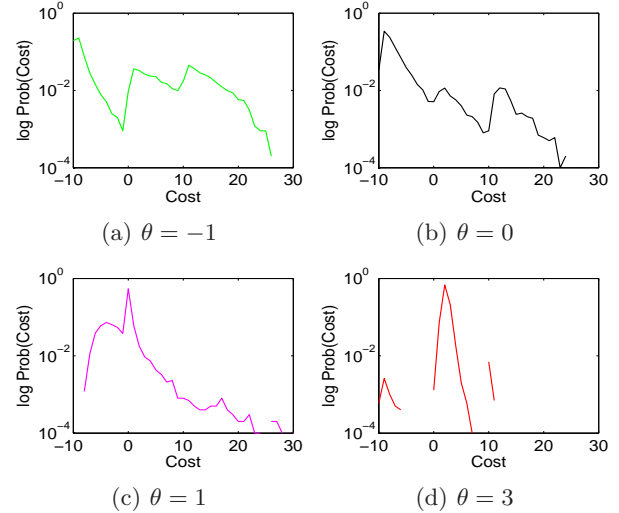

(a) $\theta = -1$

(b) $\theta = 0$

(c) $\theta = 1$

(d) $\theta = 3$

Figure 4: The log-probability of the cost in the case of a target and an adjacent threat.

Furthermore, we believe that the path integral solutions also provided insight in what is the consequence of risk sensitive control, in terms of the emerging optimistic and pessimistic behaviour. For example, in the case of both extreme optimistic or pessimistic attitude, we found apathic behaviour in the sense of zero optimal control. A nonzero $\theta$ has its effect on the distribution of the cost. In general, we found that smaller $\theta$ leads to more target oriented behaviour, while larger $\theta$ leads to more cautious behaviour, aimed to avoid threats and high costs.

From a practical point of view, risk sensitive control might be used to make control more greedy or more robust. With lower $\theta$, we found that the mode of the cost distribution decreases, however at the expense of larger tails of this distribution, i.e., more outliers with high cost. With higher $\theta$, these outliers with high costs are prevented.

The path integral method seems a promising approach, with in particular the advantage of the linearity after the log transformation. Its practical applicability is still a subject of research. One application is an algorithm for efficient reinforcement learning applied to a robot dog [19]. Another recent application is in generating animations [1, 5]. Whenever path integrals are applied for optimal control aimed to minimize expected cost, it can be applied for risk sensitive optimal control.

General conditions for existence of nontrivial solutions in risk sensitive control is an active area of research, see e.g. [4]. Providing such conditions for risk sensitive path integral control is beyond the scope of this paper and left for future research. Another possibility of future research is to explore whether the generalization or risk sensitive control is also applicable to the more general framework of KL-controls, which include path integral control as a special case [20].